\documentstyle[12pt,epsf]{article}
\newlength{\extraspace}
\newlength{\extraspaces}
\newcommand{\be}{\begin{equation}
\addtolength{\abovedisplayskip}{\extraspaces}
\addtolength{\belowdisplayskip}{\extraspaces}
\addtolength{\abovedisplayshortskip}{\extraspace}
\addtolength{\belowdisplayshortskip}{\extraspace}}
\newcommand{\ee}{\end{equation}}
 
\newcommand{\ba}{\begin{eqnarray}
\addtolength{\abovedisplayskip}{\extraspaces}
\addtolength{\belowdisplayskip}{\extraspaces}
\addtolength{\abovedisplayshortskip}{\extraspace}
\addtolength{\belowdisplayshortskip}{\extraspace}}
\newcommand{\ea}{\end{eqnarray}}

\newcommand{\bas}{\begin{eqnarray*}
\addtolength{\abovedisplayskip}{\extraspaces}
\addtolength{\belowdisplayskip}{\extraspaces}
\addtolength{\abovedisplayshortskip}{\extraspace}
\addtolength{\belowdisplayshortskip}{\extraspace}}
\newcommand{\eas}{\end{eqnarray*}}
 
\newcounter{subequation}[equation]
\makeatletter

\expandafter\let\expandafter
\reset@font\csname reset@font\endcsname

\def\subeqnarray{\arraycolsep1pt
    \def\@eqnnum\stepcounter##1{\stepcounter{subequation}%
        {\reset@font\rm(\theequation\alph{subequation})}}
\jot5mm     \eqnarray}

\makeatother

\newcommand{\NP}[1]{Nucl.\ Phys.\ {\bf #1}}
\newcommand{\PL}[1]{Phys.\ Lett.\ {\bf #1}}
\newcommand{\CMP}[1]{Comm.\ Math.\ Phys.\ {\bf #1}}

\newcommand{\bra}{\langle}
\newcommand{\ket}{\rangle}
\newcommand{\ra}{\rightarrow}

\newcommand{\rra}{\ \longrightarrow \ }

\newcommand{\nonum}{\nonumber \\[1.5mm]}
\newcommand{\sspace}{\makebox[1cm]{ }}

\newcommand{\th}{{\theta}}

\newcommand{\sh}{{\rm sh}}
\newcommand{\ch}{{\rm ch}}

\newcommand{\cL}{{\cal L}}

\newcommand{\gr}{g_{\sc r}}
\newcommand{\n}{{\sc n}}

\begin{document}
%
\begin{titlepage}
%
%
\mbox{ } 
\vspace{1.2cm}

\begin{center}
{\LARGE Parametric holomorphy?}\\[3mm] 
{\LARGE Triviality versus Duality in Sinh-Gordon}
\vspace{1.6cm}

{\large M. Niedermaier}
\\[2mm]
{\small\sl Department of Physics, 100 Allen Hall}\\
{\small \sl University of Pittsburgh}\\
{\small\sl Pittsburgh, PA 15260, U.S.A.}
\vspace{2cm}

{\bf Abstract}
\end{center}
\vspace{-5mm}

\begin{quote}
Suppose a regularised functional integral depends holomorphically 
on a parameter that receives only a finite renormalization. Can one expect 
the correlation functions to retain the analyticity in the parameter 
after removal of the cutoff(s)? We examine the issue in the Sinh-Gordon theory 
by computing the intrinsic 4-point coupling as a function of the 
Lagrangian coupling $\beta$. Drawing on the conjectured triviality
of the model in its functional integral formulation for $\beta^2 > 8\pi$,
and the weak-strong coupling duality in the bootstrap formulation 
on the other hand, we conclude that the operations: 
``Removal of the cutoff(s)'' and ``analytic continuation in $\beta$'' 
do {\em not} commute. 
\end{quote}
\vfill
\end{titlepage}


{\em 1. Introduction:}
Weak-strong coupling dualities have been a recurrent theme in
quantum field theory; see e.g.~\cite{Olive} for a review. 
Typically either the weak or the strong coupling regime admits a 
controllable series expansion, while 
the other regime is `elusive' even in a non-perturbative formulation, 
e.g.~via a lattice-regularized functional integral. Of course this is 
what makes the duality interesting in the first place, but it also 
presents a puzzle as to its precise meaning: If the duality is a
feature of an already completely defined theory, one should be 
able to prove or disprove it, -- which appears to be utopian in most cases. 
If on the other hand it should be thought of as a definition,
supplementing an incomplete computational scheme, it is hard to
see how a non-perturbative functional integral construction 
should in principle leave room for such a supplementation. 

A mathematically controllable variant of this puzzle occurs
in the Sinh-Gordon theory. In its bootstrap formulation it is
long known to possess a weak-strong coupling duality, mapping
the super-renormalizable $\beta^2 < 8\pi$ regime into the `elusive'
$\beta^2 > 8\pi$ one, where $\beta$ is the Lagrangian coupling. 
On the other hand there are good reasons to expect the theory
to be ``trivial'', i.e.~non-interacting in its functional 
integral formulation, for $\beta^2 > 8\pi$. 

Since ``triviality'' can be expressed as the vanishing of the 
intrinsic 4-point coupling $\gr$, it is natural to examine
the issue in terms of this quantity. In the case at hand $\gr = 
\gr(\beta)$ is defined by 
\be
\gr = -\frac{M^2 G(0,0,0,0)}{G(0,0)^2}\;,
\label{i1}
\ee
where $M$ is the mass gap and G are the Green functions of the 
fundamental field $\phi$. They are related to the 
Fourier transform of the truncated (connected) Euclidean correlation 
functions by 
\be
\bra \phi(k_1) \ldots \phi(k_N)\ket_T =
(2\pi)^2 \delta^{(2)}(k_1 + \ldots + k_N) G(k_1,\ldots,k_N)\;,
\label{i2}
\ee
where $k_1 + \ldots + k_N =0$ is understood in the arguments of $G$.
In \cite{us} a technique has been developed to compute $\gr$ non-perturbatively
and to high accuracy in any integrable QFT, starting from its exact 
form factors. Later we shall use this technique to compute $\gr$ in the 
(bootstrap) Sinh-Gordon model and then return to the above issue.
Let us first however lay out the problem in more detail.
\bigskip

{\em 2. Triviality versus duality:}
Classically the Sinh-Gordon model is described by the Lagrangian
$$
\cL = \frac{1}{2} (\partial_{\mu}\phi)^2 - 
\frac{M_0^2}{\beta^2}\cosh \beta \phi(x)\;,
$$
where $M_0$ is a (bare) mass parameter and $\beta$ a coupling 
constant. The QFT is presumed to describe the scattering of a 
single massive stable particle of (physical) mass $M$.   
It can be constructed both via a functional integral formulation
and in terms of the form factor bootstrap. As this will be 
relevant for the intrinsic coupling we briefly review the 
main features known or expected to hold in both approaches.

In a functional integral formulation the model can be regularized
by Wick ordering, say with respect to a UV regularized free
propagator of mass $\mu$. One can adopt a scheme in which $\beta$
is not renormalized and the model is then thought to have a
(asymptotic) super-renormalizable perturbative expansion in $\beta$,
for $\beta^2 < 8\pi$. One will require that physical quantities are invariant 
under the normal-ordering renormalization group, i.e.~that they are 
annihilated by the differential operator $\mu(\partial/\partial \mu) -
(\beta^2/8\pi)M_0(\partial/\partial M_0)$. In particular after 
removal of the cut-off all Green functions will be functions
of $\beta\,,M_0$ and the ratio $\mu/M_0$, that are invariant in 
the above sense. Moreover one can show (by so-called tadpole dressing 
\cite{DestVega,MNspec}) that the dependence on $\mu/M_0$ enters only through the 
physical mass $M$. The latter, i.e. its relation to the Lagrangian parameters    
is known exactly \cite{SinhM} 
\ba
&& M = \frac{4 \sqrt{\pi}}{\Gamma(\frac{1}{2} - \frac{B}{4})
\Gamma(1 + \frac{B}{4})}\left[ - 
\frac{\Lambda\, \Gamma(\frac{B}{2-B})}{16 \Gamma(-\frac{B}{2-B})}
\right]^{(2-B)/4}\;,\nonum
&&\mbox{where}\quad B = \frac{2\beta^2}{8\pi + \beta^2}\;,
\quad \Lambda = M_0^2 \,\mu^{\beta^2/4\pi}\,.
\label{Shmass}
\ea
Note that $\Lambda$ is renormalization group invariant and 
$\Lambda^{(2-B)/4} = M_0(\mu/M_0)^{B/2}$ produces the correct mass dimension.
As it stands the expression (\ref{Shmass}) is physically meaningful 
only for $0<\beta^2/8\pi<1$, though it happens to be real 
also for $2n <\beta^2/8\pi < 2n+1$, with $n$ a positive integer. 
The eigenvalues of the infinite set of higher conserved charges
\cite{MNspec} exhibit the same feature through their dependence 
on the exact tadpole function (which can be determined from (\ref{Shmass})). 

For $\beta^2 > 8\pi$ the status of the model defined through the functional 
integral approach is not rigorously known, however there are good reasons
to expect it to be non-interacting. In particular the intrinsic
coupling would then vanish for $\beta^2 > 8\pi$ when one attempts to remove
the cutoff(s). The rationale behind this expectation are the properties of
two closely related models, the Liouville model \cite{Liou} and the
Sine-Gordon model \cite{DimHur}, where constructive QFT techniques have
been used to derive triviality results for $\beta^2 >8 \pi$.
Specifically the methods used for the exponential interaction
\cite{Liou} can be transferred to the Sinh-Gordon case and allow to
construct the model at least for $\beta^2 < 4\pi$; for $\beta^2 > \beta_0$
with $\beta_0$ probably $8\pi$ the infared regularized model is
shown to be non-interacting when the ultraviolet cutoff is
removed. Methods of the rigorous renormalization group were
used by Dimock and Hurd \cite{DimHur} to study the Sine-Gordon model:
their results suggest triviality of that model for $\beta^2 > 8\pi$; so
far they have not been adapted to the exponential interaction or the
Sinh-Gordon model.

In the bootstrap approach on the other hand the model is meant to 
exist and to be interacting for all $0< \beta^2 < \infty$. 
In particular the intrinsic coupling, which we are going to 
compute below, is expected to be non-vanishing for all $\beta \neq 0$. 
As usual the bootstrap construction starts from the exact 
S-matrix. The S-matrix is purely elastic and is postulated to be 
\cite{SinhS} 
\be
S(\th) = \frac{\sh \th -i\sin\frac{\pi}{2} B}%
{\sh\th +i\sin\frac{\pi}{2} B}\;,
\label{Sh1}
\ee
with $B$ as in (\ref{Shmass}).
The specific dependence on the Lagrangian coupling $\beta$ has 
been tested in two-loop perturbation theory \cite{Sas}. 
The (relevant) form factors computed from this S-matrix will be 
supplied below. Both the S-matrix and the form factors are manifestly 
invariant under the `duality' transformation 
\be
\beta\rra \frac{8\pi}{\beta}\;\;\;\mbox{or}
\;\;\;B \rra 2 - B\;,
\label{Sh2}
\ee
mapping the super-renormalizable regime into the `elusive' $\beta^2 > 8\pi$
one. In principle the Green functions can be constructed from the form 
factors so that the QFT defined by the form factor bootstrap
will exhibit the duality (\ref{Sh2}), -- provided the physical mass 
is declared to be a finite invariant numerical parameter (unlike (\ref{Shmass})
which vanishes for $\beta^2/8\pi \ra 1^-$). In quantities like the intrinsic 
4-point coupling $\gr$ where the mass drops out the latter problem is absent. 
In the following we shall compute $\gr$ within the form factor bootstrap and 
find it to be duality invariant as expected
\be
\gr(B) = \gr(2 - B)\;.
\label{Sh3}
\ee
A plot of $\gr(B)$ versus $B$ in the bootstrap theory can be found 
in Fig.~1 below. 
\bigskip

{\em 3. Computation of the intrinsic coupling:} 
In principle it seems straightforward to compute $\gr$ from the 
known form factors of a theory. One simply inserts a resolution
of the identity in terms of scattering states, once for 
the two-point function and three times for the four-point function 
in (\ref{i2}). The technical problem is that a large number of 
distributional terms will appear that mask the expected analyticity 
in the momenta and which also render the expessions rather unwildly.
By a careful rearrangement however partial sums of the distributional 
terms can be decomposed into a regular term and a singular remainder 
whose coefficient vanishes identically. In particular the analyticity 
in the momenta then becomes visible and the zero momentum limit relevant 
for (\ref{i1}) can be taken. Clearly $\gr$ will decompose in that way 
into a fourfold infinite sum over the phase space of the inserted 
multi-particle states. Remarkably the dominant term, coming from the 1-particle 
form factor contribution in the two-point function and the 1+3-particle
contributions in the 4-point function, typically gives already $98\%$~(!)~of
the full answer \cite{us}. Moreover for this dominant contribution a 
general model-independent expression in terms of the derivative of the 
S-matrix and the 1-and 3-particle form factors can be obtained. The general 
formula can be found in \cite{us}; here we need only the case where the 
bootstrap S-matrix $S(\th)$ is scalar, diagonal, and without bound state poles. 
For the $n$-particle form factors of a scalar operator one can then write 
$F^{(n)} = F(\th_n,\ldots,\th_1)$,
where the rapidities $\th_j$ parameterize the on-shell momenta 
of the $n$-particle scattering state. The formula for the coupling then reads
\be
\gr = -12 i\frac{d}{d\th} S(\th)\bigg|_{\th =0}   
- 24 \int_0^{\infty} \frac{du}{4\pi} \bigg[- \frac{4}{u^2} 
+ \frac{1}{16 \ch^2 u}\,|F(i\pi,-u,u)|^2  \bigg] + \ldots \;.
\label{i4}
\ee
Here the dots indicate terms coming from $n\geq 3$-particle 
contributions to the two-point function and from the $n \geq 5$-particle
contributions to the 4-point function. Further since $\gr$ does not depend 
on the normalization of $F^{(1)}$ we normalized $F^{(3)}$ relative to 
$F^{(1)} \equiv 1$.

To evalute (\ref{i4}) we now prepare the relevant form factors.
The form factors of the Sinh-Gordon model factorize into 
a universal transcendental function (independent of the local 
operator considered) and a polynomial remainder.
The universal part is essentially a product of minimal form 
factors
\be
\psi(u) = - i {\cal C}\,\sh\frac{u}{2} 
\exp\left\{ -2 \int_0^{\infty} \frac{ds}{s} 
\frac{\ch \frac{(B-1)}{2}s}{\ch\frac{s}{2}\,\sh s}\;
\sin^2 \frac{s}{2\pi}(i\pi -u) \right\} = : 
-i\sh\frac{u}{2} \,\psi_0(u)\;.
\label{Sh4}
\ee 
$\psi(u)$ is analytic in the strip $0 < {\rm Im}\,u < \pi$, has 
a simple zero at $u =0$ and no others in the strip
of analyticity. The normalization constant ${\cal C}=\psi(i\pi)$ is real 
and is chosen such that $\psi(u) \ra 1$ for $u \ra \pm \infty$. 
As indicated it is sometimes convenient to split off the $\sh$-term 
and work with $\psi_0(u)$. It is roughly `bell-shaped', decays 
exponentially for $u \ra \pm \infty$ and obeys
\be
\psi_0(u + i\pi) \psi_0(u) = 
\frac{2 i}{\sh u + i \sin\frac{\pi}{2}B}\;.
\label{Sh5}
\ee
The form factors of the scalar field $\phi$ then are 
$F^{(1)}(\th) = 1$ and 
\be
F^{(n)}(\th) = 
\left( \frac{8 \sin \frac{\pi}{2} B}{\psi(i\pi)} \right)^{\frac{n-1}{2}}
h^{(n)}(t) \prod_{k > l} \frac{\psi(\th_{kl})}{t_k + t_l}\;,
\sspace n\;\mbox{odd},\;n\geq 3\;.
\label{Sh6}
\ee
Here $t_j = e^{\th_j},\;j = 1,\ldots, n$, and $h^{(n)}(t)$ is 
a symmetric polynomial in $t_1,\ldots,t_n$ of total degree
$n(n-1)/2$ and partial degree $n-2$. They are conveniently 
expressed in terms of elementary symmetric polynomials
$\sigma_k^{(n)}, \;k =1,\ldots,n$. For example \cite{KM}
\be
h^{(1)}(t) = 1\;,\;\;\; h^{(3)}(t) = \sigma_3\;,\;\;\;
h^{(5)}(t) = 
\sigma_5[\sigma_2 \sigma_3 - 2 \cos\frac{\pi}{2}B \,\sigma_5]\;,
\label{Sh7}
\ee
where we suppressed the superscripts $(n)$ on the right hand side.
Generally $h^{(n)}(t)$ contains $\sigma_n$ as a factor. 
In particular the 3-particle form factor more explicitly reads
\be
F^{(3)}(\th) = \frac{i\sin\frac{\pi}{2}B}{\psi(i\pi)}
\psi_0(\th_{31})\psi_0(\th_{32})\psi_0(\th_{21}) \,
{\rm th}\frac{\th_{31}}{2} {\rm th}\frac{\th_{32}}{2} 
{\rm th}\frac{\th_{21}}{2}\,.
\label{Sh8}
\ee
Inserting (\ref{Sh8}) and (\ref{Sh1}) into (\ref{i4}) one arrives at the 
result 
\be
\gr = -\frac{24}{\sin\frac{\pi}{2}B} -
\frac{3}{2\pi} \int_0^{\infty} du \left\{
\frac{{\rm th}^2 u \coth^4 \frac{u}{2}}{\ch^2 u}\,\Psi(u) 
-\frac{16}{u^2}\right\} + \ldots \;,
\label{Shgr}
\ee
where
\ba
\Psi(u) &=& \left| \frac{\psi_0(2 u)}{\psi_0(0)}\,
\frac{\psi_0(i\pi - u)}{\psi_0(i\pi)}\,
\frac{\psi_0(i\pi + u)}{\psi_0(i\pi)}\right|^2 \nonum
&=& \exp\left\{-2 \int_0^{\infty} \frac{dt}{t} 
\left(\frac{\ch \frac{(B-1)}{2}t\,\ch t}{\ch\frac{t}{2}}+
\frac{\ch {\scriptsize (B-1)}t}{\ch^2 t}\right)\;
\frac{1 - \cos\frac{2 u t}{\pi}}{\sh t} \right\}\;. 
\label{Sh13}
\ea
Fig.~1 shows a plot of $\gr(B)$ versus B, which exhibits the announced 
weak-strong coupling duality (\ref{Sh3}). For small $B$ the behavior is linear 
and the slope can be evaluated analytically%
\footnote{I thank P. Weisz for pointing this out.}
\be
\gr = (9\pi/2 + \ldots) B + O(B^2) \;.
\label{Sh14}
\ee
The dots indicate contributions to the slope coming from the subleading 
terms in (\ref{i4}). 

\begin{figure}[htb]
\hspace{4cm}
\vspace{-10mm}
\leavevmode
\epsfxsize=100mm
\epsfysize=70mm
\epsfbox{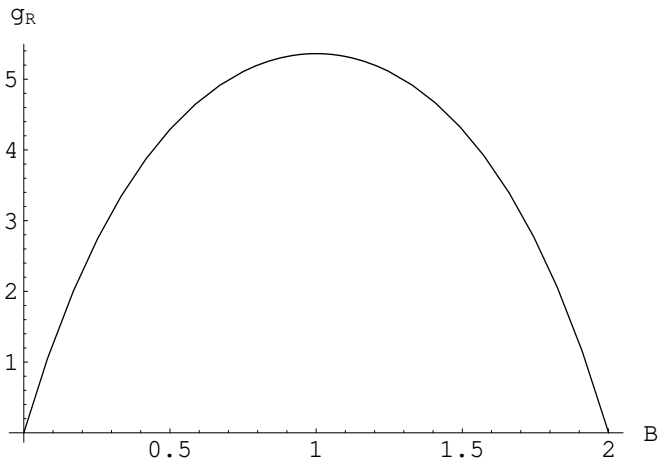}
\vspace{2mm}
\caption{Intrinsic coupling of the (bootstrap) Sinh-Gordon model} 
\end{figure}

\vspace{2mm} 
On the other hand perturbation theory (PT) 
predicts $\gr(B)_{\rm PT} = 4 \pi B + O(B^2)$, as $B = \beta^2/4 \pi 
+ O(\beta^4)$. The leading term in (\ref{i4}) thus gives only a crude approximation
to the slope. Of course the series (\ref{i4}) has been designed to rapidly
converge pointwise in $B$, so `misusing' it to extract the coefficients of 
an asymptotic expansion in $B$ one cannot expect the convergence of 
the coefficients to be equally rapid. Numerically $\gr(B)$ computed via 
(\ref{Shgr}) and $\gr(B)_{\rm PT}$ give very similar results for $0 < B< 0.1$. 
In particular a linear fit on the $0<B<0.1$ segment of Fig.~1 would give a `mock slope'
close to $4\pi$. Generally speaking this illustrates the difficulty to extract the 
coefficients of an asymptotic expansion from numerical data without having
some analytical control over the remainders. Of course next-to-leading contributions 
could in principle be computed, both in (\ref{Shgr}) and in PT, but
the principle problem would remain. 
For completeness let us also list the numerical values for $\gr(B)$ in (\ref{Shgr}) 
at a few points with $B \geq 0.1$: $\gr(0.1) = 1.270,\;\gr(0.25) = 2.725,\; 
\gr(0.5) = 4.291,\;\gr(0.75) = 5.107,\;\gr(1) = 5.362$. Guided by the experience 
with other models \cite{us} we expect them to amount to about $98\%$ of the exact 
answer. 
\bigskip

{\em 4. Removal of the cutoff(s) versus analytic continuation:}
Let us now return to the relation between the functional
integral and the bootstrap formulation of the theory. The suspected
discrepancy in the $\beta^2 > 8\pi$ behavior is most likely 
due to the fact that the operations: ``Removal of the cutoff(s) in the 
moments of the functional measure'' and ``analytic continuation in $\beta$'' 
do not commute. A useful analogy%
\footnote{due to E. Seiler (private communication).}
is the behavior of the normalized moments 
of the `Euler integral' measure, $\bra t^\n\ket_z =
\Gamma(z)^{-1} \int_0^{\infty} dt \,e^{-t} t^{z -1 + \n}$. Initially defined for 
${\rm Re}\,z >0$, they can be analytically extended to ${\rm Re}\, z <0$, 
and are entire functions: $\bra t^\n\ket_z = \prod_{j=0}^{\n-1} (z + j)$.
On the other hand the integral repesentation breaks down for 
${\rm Re}\, z<0$, but can be restored by introducing a cutoff.
The moments of the regularized measure 
$$
\bra t^\n\ket_{z,l} = \frac{1}{\Gamma(z,1/l)} \int_{1/l}^{\infty} dt \,e^{-t}
t^{z -1 + \n} \,,
$$
(i.e.~a ratio of incomplete Gamma functions) then obey
\ba
&& \lim_{l\ra \infty} \bra t^\n\ket_{z,l} = \bra t^\n \ket_z\,,
\quad \quad \,{\rm Re}\,z > 0\;,\quad\mbox{but}\nonum
&& \lim_{l\ra \infty} \bra t^\n\ket_{z,l} = 0\,,\quad \quad
\quad \,\;\;{\rm Re}\, z < 0\;. 
\label{ex}
\ea

In fact one can imitate this phenomenon in the bootstrap approach
to the Sinh-Gordon theory by describing it as the $L\ra \infty$ limit of auxiliary
continuum bootstrap models ${\rm ShG}_L$. This modified bootstrap 
construction leads to a trivial S-matrix and a vanishing
intrinsic coupling at $B=1$.%
\footnote{Note that for $B \ra 1^-$ the ingredients
of the standard bootstrap description degenerate: The S-matrix develops a pole
on the boundary of the physical strip and Zamolodchikov's mass 
function (\ref{Shmass}) (based on the thermodynamic Bethe ansatz) 
vanishes.}
To describe the auxiliary models let $B_L$ denote 
the first $L+1$ terms in a Taylor expansion of $B$ around 
$\beta^2/8\pi =0$, i.e.
\be
B_L = 2\left[ \frac{B}{2-B} - \Big(\frac{B}{2-B}\Big)^2
+ \ldots + (-)^L \Big(\frac{B}{2-B}\Big)^{L+1}\right]\;,\quad B<1\,.
\label{BL}
\ee
Let $S_L(\th)$ denote the bootstrap S-matrix (\ref{Sh1}) with 
$B$ replaced by $B_L$. This coupling dependence of the S-matrix 
would also pass an $L$-loop perturbation theory test, taking for granted 
that (\ref{Sh1}) does. In contrast to (\ref{Sh1}) however one has 
\be
\lim_{B \ra 1^-} S_L(\th) = 1\,,\quad \forall L \geq 1\;,
\quad\quad  \mbox{where} \quad S_L(\th) = 
\frac{\sh \th -i\sin\frac{\pi}{2} B_L}%
{\sh\th +i\sin\frac{\pi}{2} B_L}\;.
\label{ShL}
\ee
Let further ${\rm ShG}_L$ denote the QFT defined by the 
form factor bootstrap based on the S-matrix (\ref{ShL}).
Since the S-matrix is trivial for $B \ra 1^-$ one expects
the intrinsic coupling to vanish
\be
\lim_{B \ra 1^-}\gr(B)_{{\rm ShG}_L} = 0\,,\quad \forall L \geq 1\,.
\label{trivial}
\ee 
This is supported by the explicit computation. The intrinsic
couplings of the ${\rm ShG}_L$ models with $L=1,4,17$ are 
shown in Fig.~2.
\begin{figure}[htb]
\hspace{4cm}
\vspace{-2mm}
\leavevmode
\epsfxsize=98mm
\epsfysize=70mm
\epsfbox{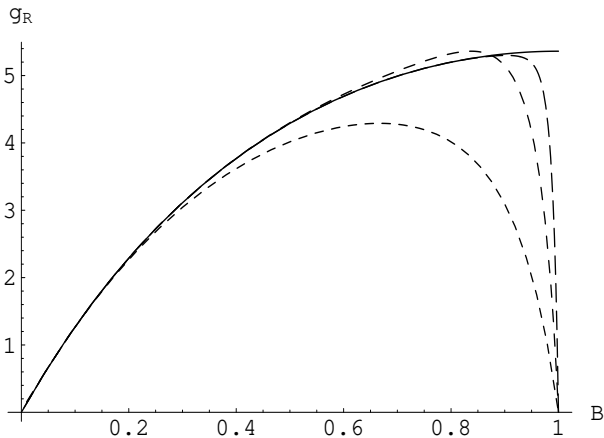}
\vspace{-3mm}
\caption{Intrinsic coupling for the ${\rm ShG}_L$ models with
$L=1,4,17$ in order of increasing slope at $B=1$. The solid line 
is the $B<1$ branch of Fig 1.} 
\end{figure}

For $B>1$ the S-matrix (\ref{ShL}) develops a pole in the physical
strip and is thus inadequate for the description of a single particle 
theory. In view of (\ref{trivial}) it is natural to extend the 
bootstrap definition of the ${\rm ShG}_L$ theory by taking $S_L(\th)=1$
for $B>1$. For large $L$ the ${\rm ShG}_L$ theory then describes a QFT
that is practically indistinguishable from the (standard bootstrap) Sinh-Gordon
theory for $B<1$, but which continuously interpolates to the trivial
theory for $B>1$. For $L \ra \infty$ the $B<1$ match with the conventional
bootstrap theory becomes exact and the transition to the trivial $B>1$
theory becomes discontinuous. The mechanism is similar to the one in the 
functional integral approach: The $L\ra \infty$ limit and analytic 
continuation in $B$ do not commute. In particular this illustrates, by way 
of an alternative, that nothing in the bootstrap formalism itself 
enforces the duality (\ref{Sh2}), (\ref{Sh3}). Strictly speaking 
it has the status of a definition.

In summary we arrive at the following scenario: Both the functional 
integral and the bootstrap approach most likely provide consistent 
non-perturbative constructions of the Sinh-Gordon theory, which 
for $\beta^2 <8\pi$ coincide. Each of the construction schemes 
suggests a natural way to define an extension into the $\beta^2 > 8\pi$
regime, which however are drastically different: Trivial in the 
former and `dual' to the $\beta^2 <8\pi$ theory in the latter case.  
The discrepancy can be understood in terms of the non-interchangeability 
of a (regularizing) limiting procedure and analytic continuation in $\beta$. 
\bigskip
\bigskip

{\tt Acknowledgement:} I wish to thank E. Seiler, P. Weisz and J. Balog 
for useful discussions on the issue and the enjoyable collaboration in \cite{us}.
The work was supported by NSF grant 97-22097.


\end{document}